\documentclass[aps,prd,twocolumn,showpacs,eqsecnum,nofootinbib]{revtex4}
\usepackage[dvips]{epsfig}
\usepackage{fancybox}
\begin{document}

\def\eV {{~\textrm{eV}}}
\def\keV {{~\textrm{keV}}}
\def\MeV {{~\textrm{MeV}}}
\def\GeV {{~\textrm{GeV}}}
\def\TeV {{~\textrm{TeV}}}
\def\PeV {{~\textrm{PeV}}}
\def\EeV {{~\textrm{EeV}}}
\def\km {{~\textrm{km}}}
\def\cm {{~\textrm{cm}}}
\def\nue {{\nu_e}}
\def\numu {{\nu_\mu}}
\def\nutau {{\nu_\tau}}   
\def\nbe {{\bar{\nu}_e}}  
\def\nbmu {{\bar{\nu}_\mu}}
\def\nbtau {{\bar{\nu}_\tau}}
\def\nubar {{\bar{\nu}}}  
\def\electron {{e^-}}   
\def\H {{\textrm{H}}}   
\def\He {{\textrm{He}}}

\title{High-energy neutrino fluxes from supermassive dark matter}

\author{\mbox{Patrick Crotty}}

\affiliation{\mbox{Department of Physics, University of Chicago, 
Chicago, Illinois 60637}
\\{\tt prcrotty@oddjob.uchicago.edu}}

\date{July 26, 2002}

\begin{abstract}
We calculate the fluxes and energy spectra of high-energy ($E > 50 \GeV$)
neutrinos from the annihilations of supermassive ($10^8 \GeV < M < 10^{16}
\GeV$), strongly interacting dark matter particles in the core of the Sun.  
We take all significant aspects of neutrino propagation through matter
into account, as well as oscillations in matter and vacuum.  We also
calculate the resulting event rates in an idealized $1 \km^3$ ice
detector.  We find that the signal should be well above background and
easily observed by next-generation neutrino detectors such as IceCube.
\end{abstract}


\maketitle


\section{Introduction}

The identity of the non-baryonic dark matter, believed to account
for about $30\%$ of the total density of the Universe
\cite{Turner:2001mw}, is one of the major unanswered questions in
cosmology today.  There is much speculation that the dark matter
consists of massive non-Standard Model elementary particles.  The
so-called ``wimps,'' or weakly interacting massive particles, are
the best known such candidate.  These are often taken as the
lightest supersymmetric particle, with a mass of no more than about
$7 \TeV$ \cite{Edsjo:1997bg}.  Any dark matter particle which is
a thermal relic cannot have a mass of more than about $340 \TeV$
\cite{Griest:1989wd}.  It has been shown that wimps could be captured by 
the Sun \cite{Press:ug,Gould:1987ir}, and that their subsequent
annihilations could produce an observable flux of high-energy neutrinos
\cite{Ritz:1987mh}.                      
    
Recently, another candidate for particle dark matter has been proposed,
the ``wimpzilla'' \cite{Chung:1998zb,Chung:1998ua,Chung:1998rq,Kolb:1998ki,Kuzmin:1998kk}.
These are gravitationally produced towards the end of inflation by the
interaction of the wimpzilla field with the inflating space-time, and in
sufficient abundance to be the dark matter.  The wimpzilla is usually
not assumed to be coupled to any other fields, although the case of
coupling between the wimpzilla and the inflaton has been studied and
wimpzilla production still found to be robust \cite{Chung:1998bt};
inflaton decays have also been shown to be a possible source of
supermassive particles \cite{Allahverdi:2001ux,Allahverdi:2002nb}.
The original wimpzilla calculations were done in the context of chaotic 
inflation ($V(\phi) = \frac{1}{2} m_\phi^2 \phi^2$), but they have 
been shown to be abundantly produced for other inflaton potentials too
\cite{Chung:2001cb}.  Wimpzillas are most efficiently produced at
extremely high masses, on the order of $10^{12} \GeV$.  

A crucial difference between wimps and wimpzillas is that the latter are
never in thermal equilibrium, and as such, their masses are not
thermodynamically constrained.  It also follows from this that their
present-day abundance does not depend on whether they have strong, weak,
electromagnetic, or only gravitational interactions (although there are
other considerations which generally rule out charged dark matter
\cite{Gould:gw,Nardi:1990ku}).

We assume in this paper that the non-baryonic dark matter consists of
wimpzillas with strong interactions, referred to from now on as
``simpzillas.''  Simpzillas, like thermal wimps, can be captured by the
Sun, and their annihilations in the solar core can produce high-energy
neutrinos. We present the results of a detailed Monte Carlo calculation
simulating the propagation of high-energy neutrinos from simpzilla
annihilations through the Sun, and also taking oscillations in matter and
vacuum into account.

Our general conclusion is that the flux of high-energy neutrinos from
simpzilla annihilations should be well above background for a broad
range of parameter space, and observable by next-generation neutrino
detectors such as IceCube.

\section{Initial neutrino flux}

\begin{figure}[t]
\centerline{\epsfxsize=3.25in \epsfbox{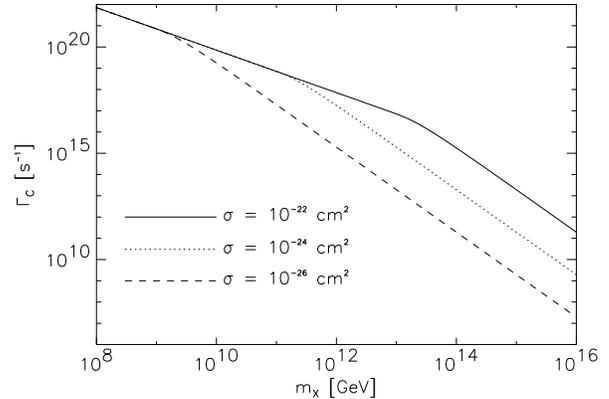}}
\caption{\label{Gamma_C_fig} The capture rate of simpzillas by
the Sun as a function of simpzilla mass $m_X$, for three different choices
of the interaction cross section $\sigma$.  In the shallower parts of
the curves, most of the simpzillas are captured.  In the steeper parts,
the simpzilla energy loss in the Sun is inefficient and only those
with lower velocities are captured.  
}
\end{figure}

Albuquerque, Hui and Kolb \cite{Albuquerque:2000rk}, henceforth AHK,
derived the capture rate of simpzillas by the Sun.  The capture of thermal 
wimps by the Sun had previously been studied in 
\cite{Press:ug,Gould:1987ir}.  AHK assumed that the dark matter consists
entirely of simpzillas in a Maxwell-Boltzmann velocity distribution,
and that the interaction cross section $\sigma$ is of the order of
the strong force, implying that the Sun is many interaction lengths thick.  
AHK's formulas for the capture rate are given in equations (2.5) and (2.7)
of their paper.  We plot the capture rate $\Gamma_C$ in Figure
\ref{Gamma_C_fig}.  The capture rate has two different forms, determined
by the efficiency of the simpzilla energy loss in the Sun.

Once captured by the Sun, the simpzillas rapidly fall to the core
and annihilate with each other.  AHK showed that equilibrium between
annihilation and capture is reached very early in the lifetime of the
Sun, and would obtain today.  In equilibrium, by definition,
$\Gamma_A = \Gamma_C/2$ (each annihilation destroys two simpzillas).
Note that this has the effect of making the equilibrium value of
$\Gamma_A$ dependent only on the interaction cross section $\sigma$
and not the annihilation cross section $\sigma_A$.  

High-energy neutrinos are produced by the simpzilla annihilations,
which produce a quark or gluon pair that then fragment into hadronic
jets containing a large number of particles.  AHK used the fragmentation
function formalism of \cite{Hill:1982iq} to calculate the numbers of
hadrons produced per annihilation.  In the dense solar core, hadrons
composed of light and charmed quarks lose most of their energy before
decaying.  Hadrons with bottom and top quarks, however, have much shorter
lifetimes and decay before substantial energy losses.  The neutrinos from
these decays are at high energies.

In this paper, as AHK did, we consider the high-energy neutrinos
from top hadrons produced in simpzilla annihilations.  Although
high-energy neutrinos can also come from bottom hadrons, it
is generally quite difficult to calculate their flux due to the
large number of $B$ meson decay modes, the products of which may
interact in the Sun before themselves decaying.  Since we find that the
neutrino flux from top hadron decays alone is well above background,
including the neutrinos from $B$ decays will only add to an already
detectable signal, although most of these will be below the $50 \GeV$
cutoff energy we choose.         

Approximately $2.8 \times 10^5 \sqrt{M_{12}}$ top hadrons are produced
per simpzilla annihilation, where $M_{12} \equiv m_X/10^{12} \GeV$.
Top quarks almost always decay in the channel $t \rightarrow
W b$.  The $W$, in turn, decays with equal branching ratios (each
about $10.5\%$) into $e \nue$, $\mu \numu$, and $\tau \nutau$.  The
$W$, like the $t$, decays before virtually any energy loss.  When the
$\tau \nutau$ pair is produced by the $W$ decay, the $\tau$ also
decays before losing much energy, producing a second $\nutau$.
About $18\%$ of the time, it also produces a second $\nue$ and another
$18\%$ of the time a second $\numu$.  

AHK showed that the top hadrons have an energy distribution proportional
to $E^{-3/2}$.  They also calculated the distribution of the subsequent
neutrinos.  The total number of neutrinos produced above $50 \GeV$
from this $t \rightarrow W$ decay chain is
\begin{equation}
\label{other_kappa_eq}
\frac{d \Phi_0^\ell}{dt} \approx \kappa_\ell 10^4 \sqrt{M_{12}} \Gamma_A
\,,
\end{equation}
  
\noindent where $\ell$ denotes the neutrino flavor ($e$, $\mu$, or
$\tau$).  $\kappa_\ell = 1$ for $\ell = \tau$ and $1/2$ for the
other two flavors, representing the fact that roughly twice as many
$\nutau$ are produced per annihilation. 

The initial neutrino energy distribution calculated by AHK is
\begin{eqnarray}
\label{initial_neutrino_flux_eq}
& & \frac{d \Phi_0^\ell}{dE dt} = \kappa_\ell 10^4 \sqrt{M_{12}} \Gamma_A
\sqrt{E_{min}} \times 0.939 \nonumber \\
& \times & \frac{E + m_W}{\sqrt{\left[ E + m_t \right] \left[ \left( E +
m_t
\right)^2 - m_t^2 \right] \left[ \left( E + m_W \right)^2 - m_W^2
\right] }} \nonumber \\
& \times & \Theta \left( E - E_{min} \right) 
\,.
\end{eqnarray}

\noindent Here, $E_{min} \equiv 50 \GeV$.  This is near the 
lower limit of $\numu$ energies detectable by IceCube
\cite{Ahrens:2001aa}, and also roughly the lowest neutrino energy
possible in the two-body $W$ decay.  

\begin{figure}[t]
\centerline{\epsfxsize=3.25in \epsfbox{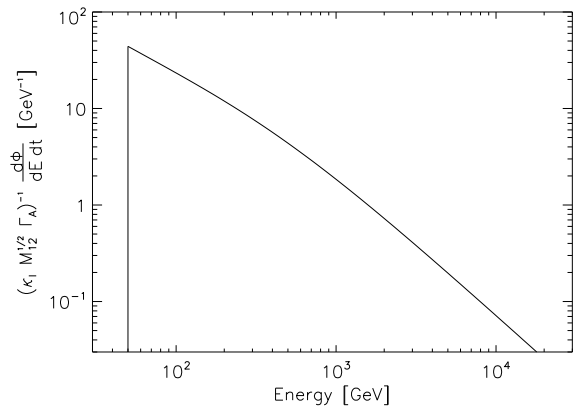}}
\caption{\label{initial_flux_fig} The initial energy distribution
of neutrinos from top hadrons produced in simpzilla annihilations.
$\kappa_\ell = 1$ for $\nutau$ and $1/2$ for $\nue$ and $\numu$.
Note that the typical neutrino energy is much lower than the
simpzilla mass, due to the large numbers of neutrinos produced
per annihilation.
}
\end{figure}

In Figure \ref{initial_flux_fig}, we show the initial simpzilla 
neutrino flux at the core of the Sun.  

\section{Neutrino interactions}

Neutrinos with energies above about $100 \GeV$ have significant
interactions with matter as they propagate through the Sun.  The
majority are charged-current (CC) interactions with nucleons.  The rest
are neutral-current (NC) scatterings; the cross sections for the latter
are about one-third of the charged-current scattering cross sections at
the same energy.  Neutrino-electron scatterings in this energy range
are negligible compared to neutrino-nucleon scatterings\footnote{Except
for $\nbe$ in the vicinity of the Glashow resonance at $6.3 \PeV$.
However, the fluxes we consider are far below this.}
\cite{Gandhi:1995tf}, and accordingly we do not consider them in our
code.  

\begin{figure}[t]
\centerline{\epsfxsize=3.25in \epsfbox{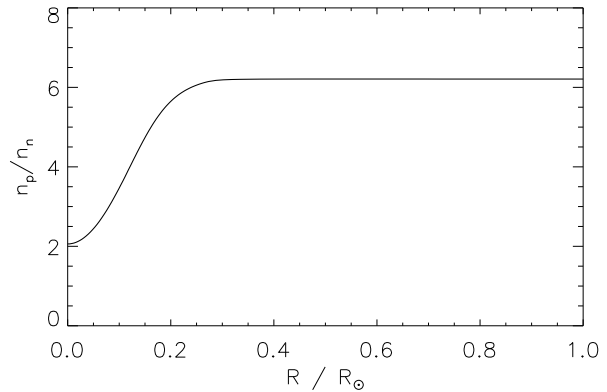}}
\caption{\label{np_ratio_fig}  The ratio of the proton and
neutron number densities in the Sun as a function of radius.
The number of neutrons is greatest in the core where more hydrogen
has been processed into helium.  Isotopes other than $^1\textrm{H}$ and
$^4\textrm{He}$ are neglected.
}
\end{figure}

We have calculated the charged and neutral-current neutrino-nucleon
cross sections between $1 \GeV$ and $10^9 \GeV$.  The most recent
previous calculations in this energy range
\cite{Gandhi:1995tf,Gandhi:1998ri} have assumed an isoscalar nucleon;
that is, one in which the quark distribution functions are the average
of those for the proton and neutron.  This is a good approximation
for a medium such as rock (the calculations were in the context of
high-energy neutrino beams propagating through the Earth).  However,
the proton number density in the Sun is between two and six times
the neutron number density, and the isoscalar approximation does not hold.
We show the ratio of proton and neutron number densities in Figure
\ref{np_ratio_fig}, calculated using the Standard Solar Model of
\cite{Bahcall:ks}.  

\begin{figure}[t]
\centerline{\epsfxsize=3.25in \epsfbox{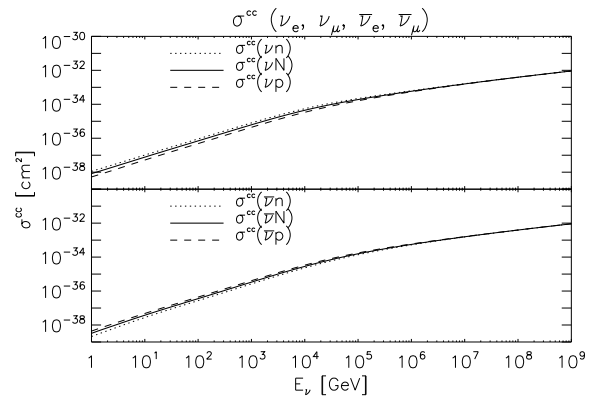}}
\caption{\label{cs_cc.e_mu_fig}  The charged-current interaction cross
section between an electron or muon neutrino and a neutron ($n$), proton
($p$), and isoscalar nucleon ($N$).
}
\end{figure}

\begin{figure}[t]
\centerline{\epsfxsize=3.25in \epsfbox{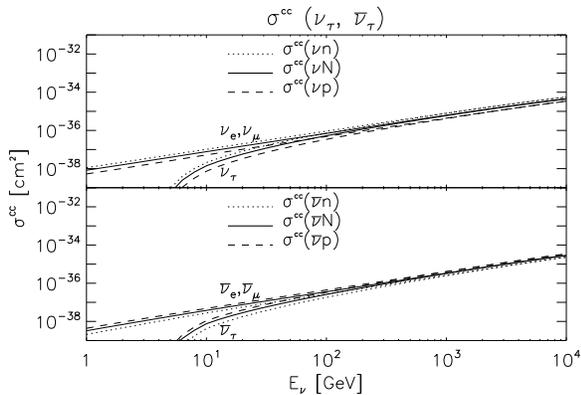}}
\caption{\label{cs_cc.tau_fig}  The charged-current cross sections
for a tau neutrino.  The electron/muon neutrino cross sections are
also shown to illustrate the kinematic suppression due to the
$\tau$ mass.
}
\end{figure}

\begin{figure}[t]
\centerline{\epsfxsize=3.25in \epsfbox{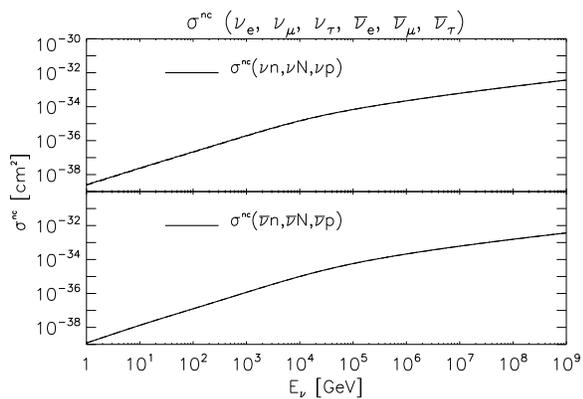}}
\caption{\label{cs_nc_fig}  The neutral-current cross sections for
all three flavors.
}
\end{figure}

We have used the recently published CTEQ6-L (leading order) parton
distribution functions \cite{Pumplin:2002vw}, together with the
deep inelastic scattering formalism of \cite{Paschos:2001np}, to
calculate the cross sections.  Given the energies involved, we may
neglect the electron and muon masses, and so the $\nue$ and $\numu$
CC cross sections are identical.  The $\tau$ mass, however, significantly
suppresses the $\nutau$ CC cross sections up to about $50 \GeV$.  
Although the effect of this on our calculation of the emergent
simpzilla neutrino fluxes is minor (we only consider energies above
$50 \GeV$, and neutrino interactions in the Sun are significant only
above about $100 \GeV$), we have taken it into account.  The NC
cross sections are identical for all three flavors.  We show our
results in Figures \ref{cs_cc.e_mu_fig}-\ref{cs_nc_fig}.

Note that for $E_\nu \lesssim 10^4 \GeV$ (and in the case of $\nutau$,
above energies where the $\tau$ mass suppression is significant),
the cross sections are approximately proportional to energy.  At
higher energies, the rate at which the cross section increases is
suppressed by the gauge boson propagator \cite{Gandhi:1995tf}.  However,
almost all of the simpzilla neutrino flux is below $10^4 \GeV$.

\begin{figure}[t]
\centerline{\epsfxsize=3.25in \epsfbox{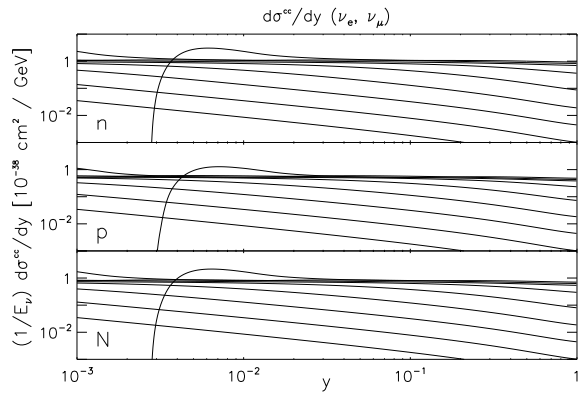}}
\caption{\label{dsigmady.nu_e.cc_fig}  CC differential cross sections for
a $\nue$ or $\numu$.  From top to bottom at $y = 5 \times 10^{-3}$,
the curves correspond to $10 \GeV$, $10^2 \GeV$, $...$, $10^9 \GeV$.  The
cutoffs in the $10 \GeV$ curves at $y \approx 3 \times 10^{-3}$ are
where the $Q^2$ of the gauge boson decreases below
$\Lambda^2_{QCD}$, and hence the parton formalism used to calculate
deep inelastic scattering becomes invalid. 
}
\end{figure}

\begin{figure}[t]
\centerline{\epsfxsize=3.25in \epsfbox{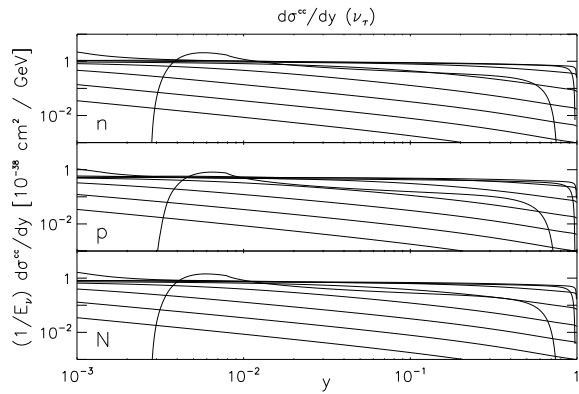}}
\caption{\label{dsigmady.nu_tau.cc_fig}  CC differential cross sections
for a $\nutau$.  The energies are the same as in Figure
\ref{dsigmady.nu_e.cc_fig}.  The cutoffs at large $y$ in the $10 \GeV$
and $100 \GeV$ curves are due to the $\tau$ mass.  
}
\end{figure}

We have also calculated the differential cross sections $d\sigma/dy$,
where the inelasticity parameter $y$ is
\begin{equation}
\label{y_def_eq}
y \equiv 1 - \frac{E'}{E_\nu}
\end{equation}

\noindent and $E'$ is the final charged lepton (neutrino) energy
in a CC (NC) scattering.  We show these for $\nue/\numu$ and
$\nutau$ CC scattering in Figures \ref{dsigmady.nu_e.cc_fig} and
\ref{dsigmady.nu_tau.cc_fig}; the others are similar.  Our code uses
these differential cross sections to obtain probability distributions
for $E'$.

We find numerically that almost all neutrino interactions take
place in the deepest part of the Sun, where the density is
greatest, out to about $0.1 R_\odot$.  This is significant
when oscillations are taken into account, because at high energies
the oscillations and interactions effectively decouple in radius (see
section V).

\section{Charged lepton interactions}

The charged leptons produced in charged-current interactions are
very important in determining the emergent neutrino flux.  Electrons
and muons tend to be stopped by their electromagnetic energy losses.
However, taus have very short lifetimes.
They tend to decay before losing any significant fraction of their
energy, and so the resulting $\nutau$ are also high-energy.  CC
interactions, in effect, absorb the $\nue$ and $\numu$ but
regenerate the $\nutau$.  This phenomenon was first remarked upon
by Ritz and Seckel \cite{Ritz:1987mh} in the context of a calculation
of neutrino spectra from thermal wimp annihilations in the Sun
(see also \cite{Kowalski:2000ri}); Halzen and Saltzberg
\cite{Halzen:1998be} have shown it to obtain for high-energy neutrino
beams propagating through the Earth (see also 
\cite{Dutta:2000jv,Becattini:2000fj,Beacom:2001xn}).

In the absence of oscillations, this means that high-energy $\nutau$
are much more likely than high-energy $\nue$ and $\numu$ to survive.
The $\nutau$ are moderated down to lower energies, to the point where the
probability of further interactions is small.  However, the total number
of $\nutau$ emerging is essentially equal to the initial number, whereas
the $\nue$ and $\numu$ are attenuated.  We will discuss the properties
of the $\nutau$ versus $\nue/\numu$ emergent fluxes in greater detail
in the penultimate section, including the effects of oscillations.

In this section, we consider the energy losses and decays of the
charged leptons in the solar medium.

\subsection{Energy losses}

As the charged leptons move through the Sun, they have electromagnetic
interactions with the medium and lose energy.  (Weak interactions,
apart from decays, are insignificant for charged leptons at
energies below $10^{16} \GeV$ \cite{Dutta:2000hh}).  The general
equation for energy loss as a function of column depth $x$ is
\begin{equation}
\label{energy_loss_eq}
-\left< \frac{dE}{dx} \right> = \alpha + \sum_j \beta^j E \,,
\end{equation}

\noindent where $\alpha$ represents interactions with electrons and 
$\{ \beta^j \}$ represent interactions with nuclei, with the
summation being over the different processes:  bremsstrahlung,
pair production, and photonuclear scattering.  Below $10^4 \GeV$,
where most of the simpzilla neutrinos (and charged leptons from
their CC scatterings) are produced, electronic losses dominate.  In
most terrestrial situations, these are ionization losses which can
be calculated with the Bethe-Bloch formula \cite{Groom:in,Groom:2001kq},
but inside the Sun the electrons are unbound.  Therefore, $\alpha$ is the
energy loss rate of a charged particle in a plasma \cite{Ritz:1987mh},
\begin{eqnarray}
\label{Ritz_Seckel_eq}
\alpha & = & \left( 9.2 \times 10^6 \right)  
\left( \frac{Z}{A \beta^2 c} \right) \nonumber \\
& & \times \left[ \ln{(2 m_e \gamma \beta^2)}
- \ln{ \left( \sqrt{ \frac{4 \pi \alpha n_e}{m_e} } \right) } \right] \,.
\end{eqnarray}

\noindent The units of $\alpha$ in Eq. (\ref{Ritz_Seckel_eq}) are 
$\textrm{GeV}/(\textrm{g}/\textrm{cm}^2)$.  The quantities $Z$ and $A$ are
the atomic number and weight of the nuclei in the medium -- for the Sun,
which is (approximately) a mixture of $^1\textrm{H}$ and $^4\textrm{He}$,
we calculate the energy losses separately for the two isotopes and
then weight them by their mass fractions.  The $\beta$ and $\gamma$ are
the Lorentz parameters of the charged lepton and $n_e$ is the electron
number density.

The nuclear energy losses become dominant at energies above about
$10^4 \GeV$, although as mentioned they are not very important for
the simpzilla neutrino flux which tends to be at energies below this.
The formulas for $\{ \beta^j \}$ are given, for example, in
\cite{Dutta:2000hh}.

\begin{figure}[t]
\centerline{\epsfxsize=3.25in \epsfbox{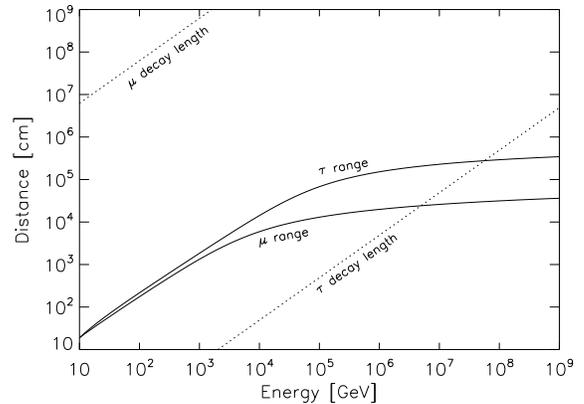}}
\caption{\label{eloss_fig}  Decay lengths and ranges of $\mu$ and
$\tau$ in the solar core (where most charged-current interactions
take place).  Note that except at the highest energies, the $\tau$
decay length is much shorter than its range; hence it decays before
losing virtually any energy.  The opposite is true for the $\mu$,
which is essentially brought to rest before decaying.
}
\end{figure}

Our code simulates both electronic and nuclear energy losses for
$\mu$ and $\tau$.  We do not follow $e$ since they never decay.  In
Figure \ref{eloss_fig}, we show the muon and tau ranges and decay
lengths in the core of the Sun, which is where the majority of 
neutrino interactions take place.  The range is defined as the
distance, in the absence of decays, over which the lepton would lose all
its energy; and similarly the decay length is the (mean) distance, in the
absence of energy losses, the lepton would travel before decaying.

\subsection{Decays}

Both muons and taus decay back into neutrinos, but the muons have
lost almost all their energy by this point and essentially decay at rest.
The resulting $\numu$ are below $50 \MeV$ and do not concern us.  The
taus, on the other hand, have negligible energy losses except at
very high energies ($E \gtrsim 10^7 \GeV$), which again does not concern
us since very little of the initial simpzilla neutrino flux is in this
range.  The $\nutau$ from the decay carries on average about $2/5$
the energy of the $\tau$, or about $1/4$ the energy of the incident
$\nutau$ (the mean energy of the $\tau$ relative to the incident $\nutau$
can be obtained from the inelasticity parameter distributions; in the
energy range we consider, the mean value of $y$ is about $0.4$, higher
than for the energies at which neutrinos interact significantly
in the Earth).  

One phenomenon our code takes into account is the production of
secondary $\nue$ and $\numu$ in $\tau$ decays, as discussed by
\cite{Beacom:2001xn}.  Approximately $18\%$ of $\tau$ decays produce
a $\numu$ (for $\tau^\mp$, a $\nbmu$ ($\numu$)), and another $18\%$
produce a $\nue$.  These secondary neutrinos are created with about
half the energy of the accompanying $\nutau$, or about $1/8$ the energy of
the incident $\nutau$.  When the $\nutau$ experiences its last few
charged-current scatterings and is downscattered to energies at which
it escapes the Sun, these secondary neutrinos emerge too.
Ref. \cite{Beacom:2001xn} considered high-energy neutrino beams from
extragalactic point sources propagating through the Earth, and showed
that the additional flux of secondaries could substantially modify
the detected signal.   

\begin{figure}[t]
\centerline{\epsfxsize=3.25in \epsfbox{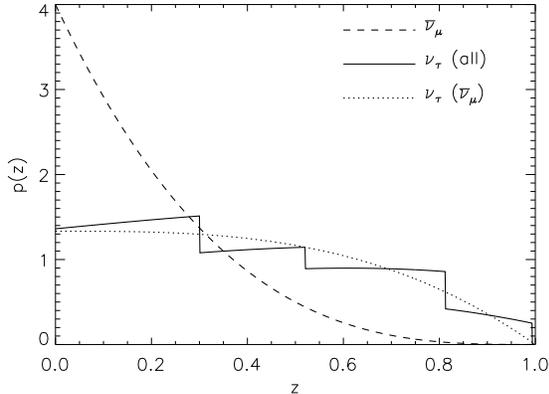}}
\caption{\label{dndz_fig}  Distributions of neutrino energies from
$\tau$ decays, with $z \equiv E_\nu/E_\tau$ and the $\tau$ assumed
to be ultrarelativistic.  Solid curve: $\nutau$
energy distribution, summing over all significant decay channels of the
$\tau$ and weighting by the branching ratios.  Some of the channels
produce massive hadrons and become kinematically impossible above a
certain $z$, the reason for the jagged appearance.  Dashed curve:
secondary $\numu$ and $\nue$ decay distributions from the $18\%$ of
$\tau$ decays which produce them.  Dotted curve:  the $\nutau$ energy
distribution in the secondary-producing channels.  We assume the
$\tau$ to have the parity it would have
if produced in a $\nutau$ CC interaction (for $\tau^-$, negative; for
$\tau^+$, positive).  
}
\end{figure}

The major decay channels of $\tau$ are $\tau \rightarrow \nutau
\mu \numu$ (18\%); $\tau \rightarrow \nutau e \nue$ (18\%);
$\tau \rightarrow \nutau \pi$ (12\%); $\tau \rightarrow \nutau \rho$
(26\%); $\tau \rightarrow \nutau a_1$ (13\%); and $\tau \rightarrow
\nutau X$ (13\%), where $X$ indicates other massive hadrons.  The
combined $\nutau$ decay distribution is given in \cite{Dutta:2000hh}.
We show it in Figure \ref{dndz_fig}, along with that of the secondary
$\nue$ and $\numu$, which is straightforward to obtain (see e.g.
\cite{Renton:td}).  Our code randomly samples over the appropriate
distributions to obtain the energies of the neutrinos from $\tau$ decays.

\section{Neutrino oscillations}

Both matter and vacuum oscillations are significant for simpzilla
neutrinos, and have a large effect on the emergent flux.  Our code
takes the oscillations of all three flavors into account.  We use
values for the mass-squared differences and mixing angles which are
consistent with recent observations.  From the SuperKamiokande
results \cite{Fukuda:1998mi,Fukuda:2000np}, we choose
$\Delta m_{31}^2 = 3 \times 10^{-3} \eV^2$, $\sin^2{\xi} = 0.1$ (which
is the upper limit on $\xi$), and $\sin^2{\theta} = 0.5$, where the angles
in the three-flavor mixing matrix are as in Eq. (3.1) of
\cite{deGouvea:2000un}.  The other parameters are constrained by solar
neutrino observations, such as those recently recorded by the Sudbury
Neutrino Observatory (SNO) \cite{Klein:2001qs,Ahmad:2002jz,Ahmad:2002ka}.
Analysis of the SNO data favors the ``large mixing angle'' (LMA) solution 
\cite{Krastev:2001tv}, and so we choose values characteristic of this:
$\Delta m_{21}^2 = 2 \times 10^{-5} \eV^2$ and $\sin^2{\omega} =
0.2$.  We assume the normal mass hierarchy with $m_1 < m_2 < m_3$,
and set the CP-violating phase in the mixing matrix to $0$.  

To simulate the oscillations, we evolve the flavor amplitudes using
the analytical solution for the time-evolution operator in
\cite{Ohlsson:1999xb,Ohlsson:2001et}.  Our Monte Carlo step sizes
are sufficiently small that the oscillation probabilities do not
change significantly over the step, and the density at each step
may be taken as constant.  If the neutrino is determined to have
had a charged-current interaction during the step, we sample
over the current probabilities to find the flavor of the resulting
charged lepton; if the lepton subsequently decays at high energy,
the resulting neutrino begins purely in that flavor.

\begin{figure}[t]
\centerline{\epsfxsize=3.25in \epsfbox{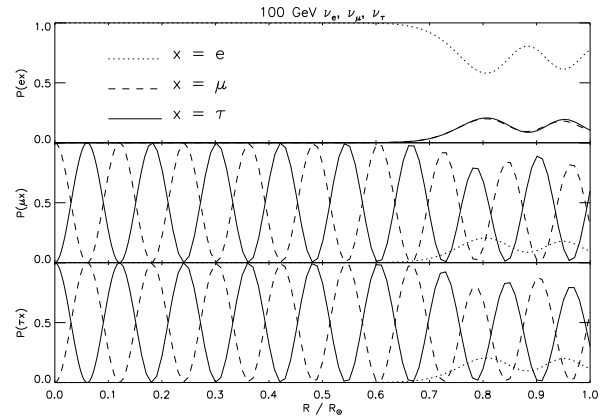}}
\caption{\label{osc.nu.100_fig}
Oscillation of $100 \GeV$ neutrinos propagating (without interactions)
from the center of the Sun to its surface.  From top to bottom:
oscillation probabilities of $\nue$, $\numu$, and $\nutau$.  Note that
the oscillations of $\nue$ are almost completely suppressed until the
resonance.
}
\end{figure}

\begin{figure}[t]
\centerline{\epsfxsize=3.25in \epsfbox{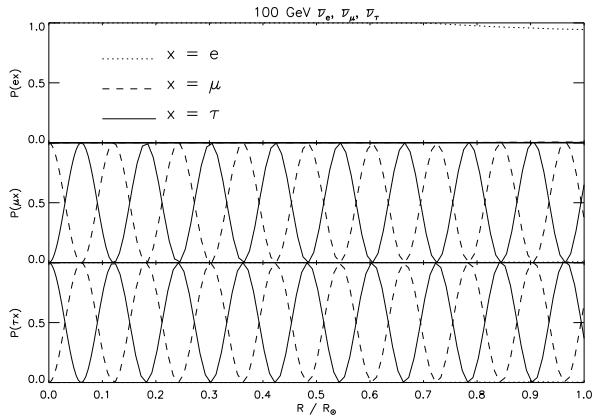}}
\caption{\label{osc.nubar.100_fig}
Oscillations of $100 \GeV$ antineutrinos propagating from the
center of the Sun to its surface.  The $\nbe$, unlike the $\nue$,
does not encounter a resonance, and its oscillations are
minimal.
}
\end{figure}

In Figures \ref{osc.nu.100_fig} and \ref{osc.nubar.100_fig}, we show
the oscillation probabilities of $100 \GeV$ neutrinos and antineutrinos
traveling through the Sun without interacting (so their energies
do not change).  This is roughly the lower limit of where neutrino
interactions in the Sun are significant.    

We have noted above that most neutrino interactions take place between
the center of the Sun and $0.1 R_\odot$.  At higher energies, the 
oscillation wavelengths increase, to the point where the oscillation
probabilities do not vary substantially over this distance.  
Interactions at energies much greater than $100 \GeV$ thus effectively
``decouple'' in radius from oscillations (though if our calculation were
extended to include sterile neutrinos, this might not hold; see
\cite{Naumov:2001ci}).

We also would not expect different reasonable choices of the oscillation
parameters to make a difference of more than a factor of two or
so in the detection rates.  Real-world detectors most easily observe
the muons created in $\numu$ and $\nbmu$ charged-current interactions
(the $\nue$ and $\nutau$ signals are much less directional),
so the relevant issue is how much different parameter choices would
change the emergent $\numu + \nbmu$ flux.  As can be seen from
Figures \ref{osc.nu.100_fig} and \ref{osc.nubar.100_fig}, the 
worst-case scenario would completely deplete the $\numu$ flux by
maximizing the $\nue$ conversion probabilities; however, the $\nbmu$ flux
would be unaffected since the $\nbe$ do not encounter a resonance and
barely oscillate.  (A difference in the sign of $\Delta m_{31}^2$,
which is currently unknown and which we assume to be positive, would
cause the $\nbe$ rather than the $\nue$ to have a resonance, so the
graphs would essentially be interchanged.)    

The $\numu \leftrightarrow \nutau$ oscillations in the Sun are
vacuum oscillations with the atmospheric mixing parameters.  The $\nue$
resonance is governed by the mixing angle $\xi$ (corresponding to
$U_{e3}$) and $\Delta m_{31}^2$.  This is in contrast to normal solar
neutrinos, whose MSW resonance is governed by $\Delta m_{21}^2$ and
$\omega$.  For the simpzilla neutrinos, that resonance is in the outermost
layers of the Sun at very low densities, and is not numerically 
significant compared to the other resonance.  We note that vacuum
oscillations outside the Sun are completely averaged out by the detector
energy resolution (see section VI).  We also note that although the
detector is considered to be looking at upcoming simpzilla neutrinos which
have come through the Earth, the density of the Earth is too high for MSW
oscillations inside it to be important.

\section{Neutrino flux at Earth}

Using the initial neutrino energy distribution given in
(\ref{initial_neutrino_flux_eq}) and taking all the physical
effects discussed above into account, we have calculated the
resulting neutrino flux at Earth.  We have also, with IceCube
in mind, calculated the event rates it would produce in an idealized 
$1 \km^3$ ice detector.

\subsection{Monte Carlo code}

We have written a Monte Carlo code to simulate the propagation
of high energy neutrinos through matter.  Each neutrino is followed
as it passes through the Sun.  The step sizes are chosen to be much
smaller than both the interaction length and the neutrino oscillation
length.  

At each step, we use a pseudo-random number generator to determine
whether an interaction occurred over that step, whether it was a
charged- or neutral-current interaction, and whether it occurred on
a proton or neutron.  In the event of a neutral-current interaction,
the final neutrino energy is determined by randomly sampling over
the appropriate inelasticity parameter distribution (see section
III).

Charged-current interactions are effectively measurements of
the neutrino flavor.  As such, when a charged-current interaction
occurs, we randomly sample over the current oscillation probabilities
to determine the flavor of the neutrino and outgoing lepton.  When
and if the lepton decays back into a neutrino, it starts in a
pure flavor state.  We calculate the initial energy of the lepton
using the inelasticity parameter distribution.  

The $\mu$ and $\tau$ produced in charged-current interactions
are followed until they decay or their energies fall below $50 \GeV$.
The $e$ produced in $\nue$ charged-current interactions are not
followed since they never decay.  (In practice, the $\mu$ almost
never decay at high energies either, but we follow them for
consistency.)  We calculate their energy losses over each step according
to (\ref{energy_loss_eq}).  We also use their boosted decay lengths
to randomly determine whether a decay occurs on each step.  If so,
we sample over the appropriate decay distribution to determine the
energy of the outgoing neutrino.  We also calculate the energies
of the secondary $\nue$ and $\numu$ in $\tau$ decays using the
energy distributions for those modes. 

Neutrino oscillations are simulated both in the Sun and in the
vacuum between the Sun and the Earth.  We average the oscillation
probabilities over $10\%$ of the Sun-Earth distance, which is
mathematically equivalent to a $10\%$ energy resolution in the
detector (characteristic of real-world detectors).

The detector is considered to be able to detect every neutrino
charged-current interaction above $50 \GeV$ taking place in a
cubic kilometer of ice.  In practice, real detectors are considerably
more complex than this and a sophisticated treatment would require
a detector Monte Carlo, which is beyond the scope of this work.  The
detector is also assumed to have an angular resolution of $1^\circ$, which
is approximately the size of the Sun on the sky.

Having calculated the neutrino flux at the Earth, we calculate the
detection rate by integrating the product of the flux, the area
of the detector ($1 \km^2$), and the probability of the neutrino
having a charged-current interaction, the integration being over
the energy $E$.  The probability of a charged-current interaction
in the detector is
\begin{equation}
P_{cc}(E) = 1 - e^{-\lambda/L_{int}(E)}
\,,
\end{equation}

\noindent with the interaction length
\begin{equation}                                                
L_{int}(E) = \frac{1}{n_p \sigma_p^{cc}(E) + n_n \sigma_n^{cc}(E)}
\end{equation}

\noindent and $\lambda = 1 \km$ is the size of the detector.
The number densities $n_n$ and $n_p$ are for ice, and the
charged-current cross sections $\sigma^{cc}(E)$ were discussed
in section III.

The background flux from atmospheric neutrinos above $50 \GeV$
is approximately \cite{Albuquerque:2001jh}
\begin{eqnarray}
\label{atmospheric_nu_flux_eq}
\frac{d \Phi_{atm}}{dE \, dA \, dt \, d\omega} =
(1.1 \times 10^{12}) E^{-3.2} \nonumber \\
\left[ \textrm{km}^{-2} ~ \textrm{yr}^{-1}
~ \textrm{deg}^{-2} \GeV^{-1} \right]
\,.
\end{eqnarray}

\noindent Integrating this over energy, the detection probability, the
detector area, and the angular size of the Sun gives about 2 
atmospheric neutrino events per year above $50 \GeV$ which come from the
direction of the Sun.  These will be almost exclusively muon neutrinos,
if we assume the detector is looking at upcoming neutrinos.  We take
this as our background.

We also note that the Baksan neutrino telescope has placed a
$90\%$ confidence level limit of muon fluxes from nonatmospheric
neutrinos coming from the direction of the Sun of about
$10^4 \km^{-2} ~\textrm{yr}^{-1}$ \cite{Boliev:xz}.  We take this
as a rough upper limit.  The future analysis of the data from
AMANDA \cite{Wischnewski:yt} may lower this limit.

\subsection{Results}

\begin{figure}[t]
\centerline{\epsfxsize=3.25in \epsfbox{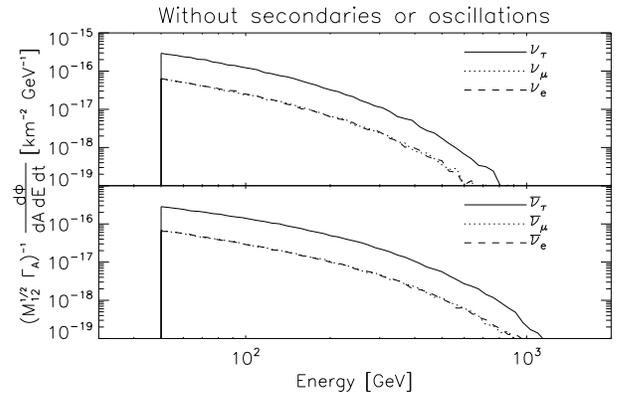}}
\caption{\label{flux.no_sec.no_osc_fig} The simpzilla neutrino
fluxes at the Earth, neglecting both neutrino oscillations and
secondary neutrinos from $\tau$ decays.  
}
\end{figure}

\begin{figure}[t]
\centerline{\epsfxsize=3.25in \epsfbox{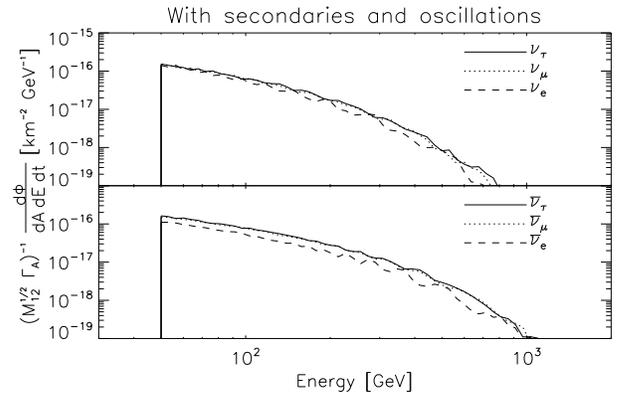}}
\caption{\label{flux.sec.osc_fig} The simpzilla neutrino fluxes
at the Earth calculated with our Monte Carlo, taking both oscillations and
secondaries into account.
}
\end{figure}

\begin{figure}
\centerline{\epsfxsize=3.25in \epsfbox{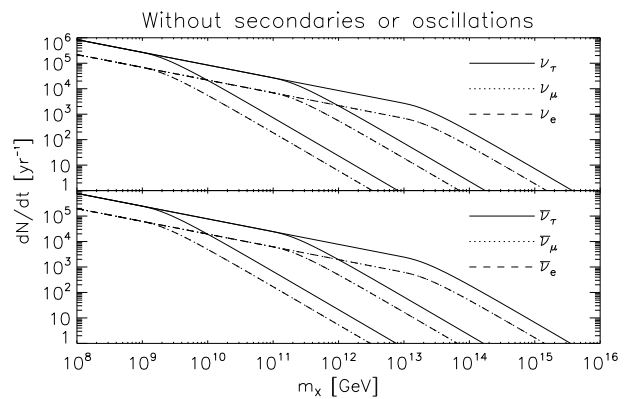}}
\caption{\label{event_rate.no_sec.no_osc_fig}  The simpzilla
neutrino event rates in a $1 \km^3$ idealized detector, neglecting
secondaries and oscillations.  The rate is plotted as a function of
the simpzilla mass $m_X$.  We show three different values for
$\sigma$, the simpzilla interaction cross section:  from left to
right, these are $\sigma = 10^{-26} \cm^2$, $10^{-24} \cm^2$, and
$10^{-22} \cm^2$.  The background from atmospheric neutrinos
($\numu$ and $\nbmu$) is approximately $2 ~\textrm{yr}^{-1}$. 
An upper limit from Baksan is about $10^4 ~\textrm{yr}^{-1}$.
}
\end{figure}

\begin{figure}
\centerline{\epsfxsize=3.25in \epsfbox{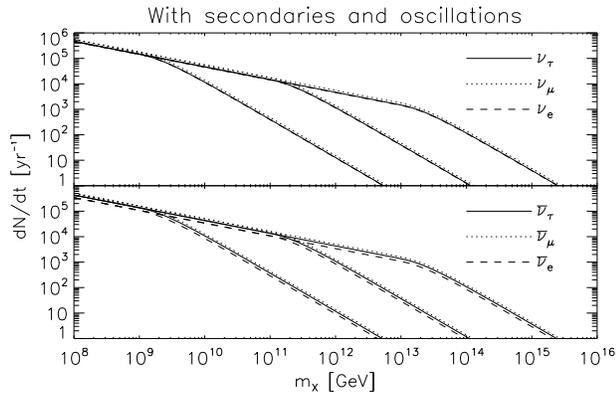}}
\caption{\label{event_rate.sec.osc_fig} The simpzilla neutrino
event rates, taking both secondaries and oscillations into account.
The same choices for $\sigma$ are plotted as in Figure 
\ref{event_rate.no_sec.no_osc_fig}.
}
\end{figure}

In Figures \ref{flux.no_sec.no_osc_fig} and \ref{flux.sec.osc_fig},
we show our numerical results for the simpzilla neutrino flux at the
Earth.  In Figures \ref{event_rate.no_sec.no_osc_fig} and
\ref{event_rate.sec.osc_fig}, we show the event rates as a
function of simpzilla mass $m_X$ for three different choices of
the interaction cross section $\sigma$.  We show two cases:  when
neither oscillations nor secondaries are included, and when both
are included.  We have found that the secondaries alone do not
make a dramatic difference to the flux, largely because it falls
over several decades in energy, and much of it is in a region
where neutrino interactions are not significant and thus secondaries
are not produced.  We accordingly do not show the other two permutations.
Previously, only the case without oscillations or secondaries has
been studied \cite{Albuquerque:2000rk,Albuquerque:2002bj}.

We discuss this case briefly.  As can be seen from
Figure \ref{flux.no_sec.no_osc_fig}, the emergent $\nutau$ fluxes
exceed those of the other two flavors by more than the initial factor
of $2$ (see section II).  This is due to the fact that the $\nutau$
which experience charged-current interactions are not, as are the
$\nue$ and $\numu$, absorbed, but rather regenerated at lower energies,
as remarked upon in section IV.

The unscattered components of the $\nue$, $\numu$, and $\nutau$ beams
relate to their initial fluxes as \cite{Albuquerque:2000rk}
\begin{equation}
\label{unscattered_flux_eq}
\frac{d \Phi_f}{dE dt} = \frac{d \Phi_0}{dE dt}
\, \, e^{-E / E_k}
\,,
\end{equation}

\noindent where $E_k$ is the ``transparency energy,'' defined as
the energy such that the mean number of interactions experienced
by the neutrino is 1.  Equation (\ref{unscattered_flux_eq}) obtains
because the cross sections are approximately linear functions of
energy in the range we are considering.  Numerically, we find
that $E_k \approx 130 \GeV$ for $\nue$ and $\numu$, $160 \GeV$ for 
$\nutau$, $200 \GeV$ for $\nbe$ and $\nbmu$, and $230 \GeV$ for
$\nbtau$.  We use (\ref{unscattered_flux_eq}) as the emergent
flux for the electron and muon neutrinos\footnote{A small number of
$\nue$ and $\numu$ (as well as $\nutau$) experience repeated
neutral-current scatterings without any charged-current scatterings.
We neglect this neutral-current-scattered component here, although
it is taken into account by our code.}.

\begin{figure}[t]
\centerline{\epsfxsize=3.25in \epsfbox{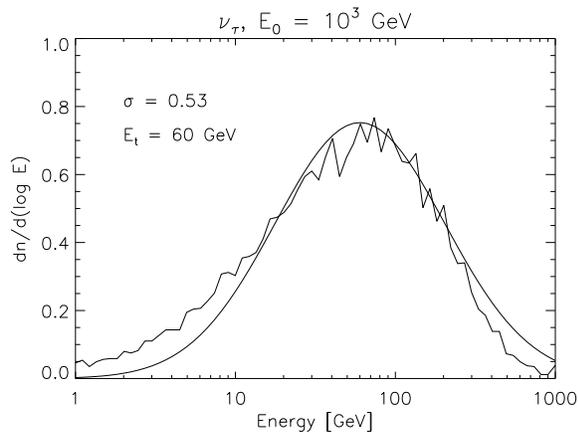}}
\caption{\label{nu_tau.log_normal.fixed_fig} Distribution of
emergent $\nutau$ energies for a monoenergetic beam with initial
energy $E_0 = 10^3 \GeV$.  We have calculated the emergent distribution
for other choices of $E_0$ too, and found it to be generally independent
of $E_0$.  We plot as a distribution in $\log{E}$
rather than $E$, and neglect oscillations.  The distribution corresponds
approximately (although obviously not exactly) to the log-normal
distribution with the parameters shown.
}
\end{figure}

To model the scattered component of the $\nutau$, we use the
observation of ourselves and several other groups that the scattered
$\nutau$ emerge in a roughly log-normal distribution.  We find
numerically that about $80\%$ of the emergent $\nutau$ and $\nbtau$
are scattered.  The log-normal distribution is defined as
\begin{equation}
\label{log_normal_dist_eq}
\frac{dn}{dE} = \frac{1}{\sqrt{2 \pi} \ln{10} \, \sigma E}
\exp{\left[ -\frac{1}{2 \sigma^2} \log^2{\left(\frac{E}{E_t}\right)}
\right]}
\,,
\end{equation}

\noindent and when transformed to a distribution in $\log{E}$, is
a Gaussian with mean $\log{E_t}$ and standard deviation $\sigma$
(whose units are decades).  Note that $E_t \ne E_k$, although they turn
out to be similar.  We show how the numerical results compare to a
log-normal distribution in Figure \ref{nu_tau.log_normal.fixed_fig}.  The
correspondence is not exact, and the log-normal fit overestimates the
high-energy neutrinos while underestimating the low-energy ones.  However,
it is evidently a decent approximation.  The $\nutau$ log-normal
fit has $\sigma = 0.53$ and $E_t = 60 \GeV$; the $\nbtau$ has
$\sigma = 0.49$ and $E_t = 113 \GeV$.  

\begin{figure}[t]
\centerline{\epsfxsize=3.25in \epsfbox{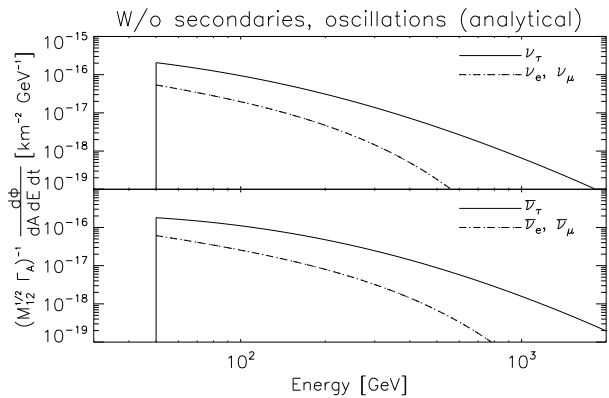}}
\caption{\label{flux.an.no_sec.no_osc_fig} Analytical approximation
for the emergent simpzilla neutrino fluxes, neglecting secondaries
and oscillations.  We assume the $\nue$ and $\numu$ fluxes to
consist entirely of the remaining unscattered neutrinos, and the
$\nutau$ flux to be the sum of an unscattered component and a
scattered component obeying a log-normal distribution.  We make
similar assumptions about the antineutrinos.  The parameters
used are given in the text.    
}
\end{figure}

In Figure \ref{flux.an.no_sec.no_osc_fig}, we plot these analytical
approximations.  The $\nue$ and $\numu$ fluxes are in qualitative
agreement with the numerical results.  The $\nutau$ fluxes are 
somewhat lower at low energies and higher at high energies than
are the numerical results, again due to the log-normal underestimating
the first while overestimating the second.  We note that Eq. 
(\ref{log_normal_dist_eq}) differs from Eq. (4.6) of 
\cite{Albuquerque:2000rk}, leading to a much smaller difference
between the $\nutau$ and $\nue/\numu$ event rates; this is due 
to our correctly taking the $1/E$ term in the Jacobian factor in
the log-normal distribution into account.

The oscillations clearly have a substantial effect on the flux.  The
$\nutau$ are depleted, whereas the $\numu$ and $\nue$ are enhanced.  In a
real world detector, this is advantageous, because it is generally easiest
to observe muons from $\numu$ charged-current interactions.  The
detectability of the simpzilla neutrino flux is therefore improved by
oscillations.

\begin{figure}[h]
\centerline{\epsfxsize=3.25in \epsfbox{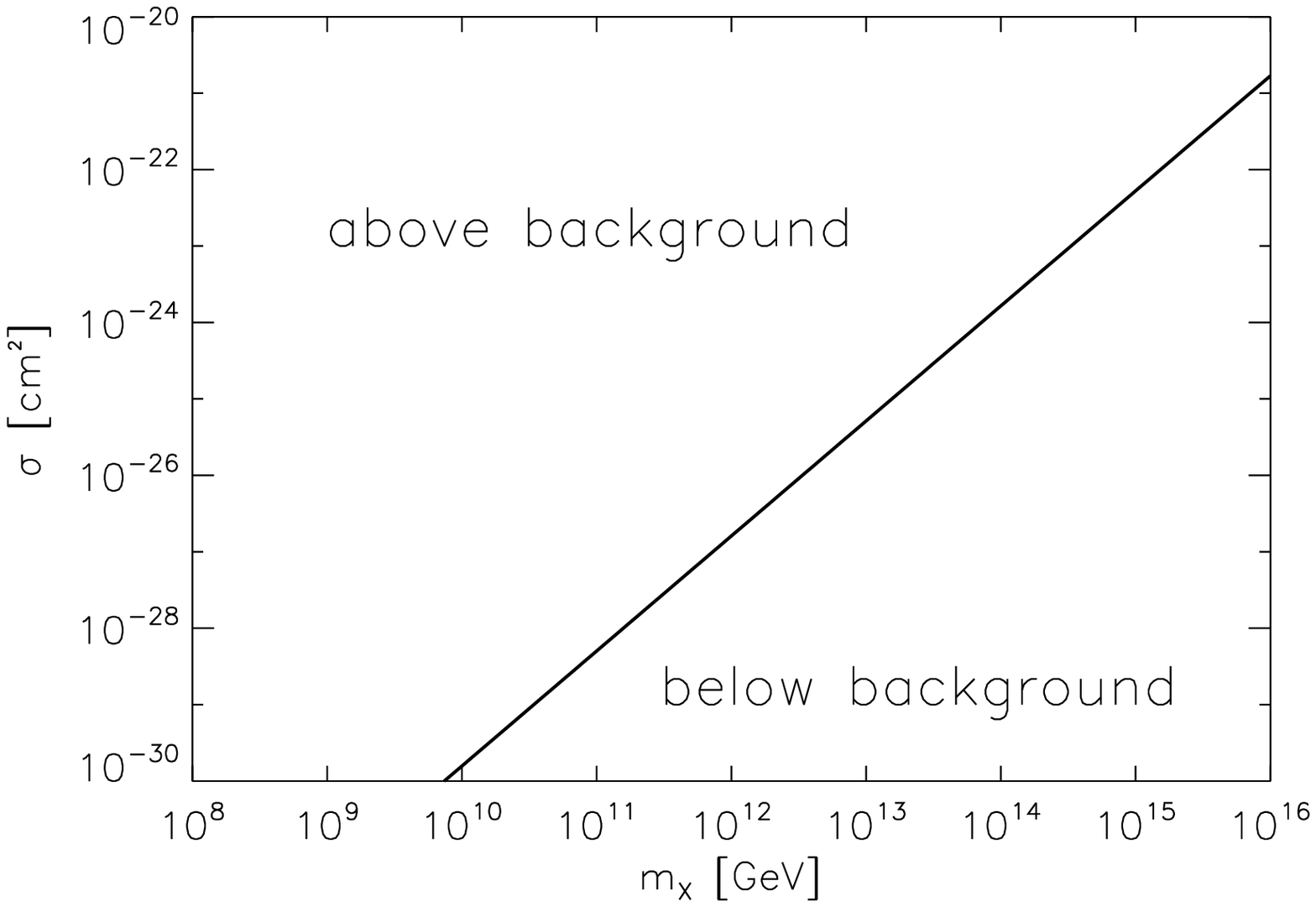}}
\caption{\label{background_threshold_fig} The regions of simpzilla
parameter space which produce neutrino fluxes (in each flavor) that are
above and below the background event rate of $2 ~\textrm{yr}^{-1}$.
If no high-energy neutrino signal is observed from the Sun, it would
definitively exclude the range of simpzilla mass and cross section
to the left of the line.  The plot was generated using the event
rates which include both secondaries and oscillations.
}
\end{figure}

In Figure \ref{background_threshold_fig}, we show the excludable
region of simpzilla parameter space; that is, the range of $m_X$
and $\sigma$ which would be ruled out by the failure to observe
a high-energy neutrino signal coming from the direction of the Sun.
Even better, of course, would be the observation of such a signal.
However, a more detailed analysis, which we leave to others, would be
necessary to rule out other possible sources such as thermal wimps, 
particularly massive ones that could produce neutrinos in roughly the same 
energy range.

\section{Conclusions}          

We have shown that the annihilations of simpzillas in the Sun should
produce a conspicuous high-energy neutrino signal in a $1 \km^3$
ice detector for a fairly large range of simpzilla mass and interaction
cross section.  Much of this range of parameter space has not yet been
ruled out by current observations.     

Our flux calculation takes all important physical effects into account,
including neutrino oscillations, which approximately equalize the fluxes
of the different flavors.  In calculating the event rates, we have
made some simplifying assumptions about the detector, and a more 
realistic calculation (say, for IceCube) would take the actual detector
response characteristics into account.  However, we would not expect this
to alter our general conclusion that the simpzilla neutrino signal would
be observable in such a detector.

The next generation of neutrino detectors should be able to either rule
out a large range of simpzilla parameter space, or provide observational
evidence for them. 

\section*{ACKNOWLEDGMENTS}

I thank E. W. Kolb for his guidance and support.  I also thank J. F.
Beacom for assisting me in a large number of ways, and pointing out
the phenomenon of neutrino interaction-oscillation decoupling in
radius.  I thank A. B. Balantekin, A. de Gouv\^{e}a, D. E. Groom, D.
M\"{u}ller, T. Ohlsson, D. Rainwater, J. Truran, M. Turner, and T. Witten 
for useful discussions.

I was supported by DOE grant No. 5-90098 while working on this project.
   


\end{document}